\documentclass[superscriptaddress]{revtex4}

\usepackage{epsf}
\usepackage{amssymb}
\newcommand{\be}{\begin{equation}}
\newcommand{\ee}{\end{equation}}
\newcommand{\bea}{\begin{eqnarray}}
\newcommand{\eea}{\end{eqnarray}}

\newcommand{\im}{i}

\begin{document}

\title{Information Flow through a Chaotic Channel: Prediction and
  Postdiction at Finite Resolution}
\author{Richard Metzler}
\affiliation{New England Complex Systems Institute, 
 24 Mt. Auburn St., Cambridge, MA 02138, USA}
\affiliation{Department of Physics, Massachusetts Institute of Technology,
 Cambridge, MA 02139, USA}
\author{Yaneer Bar-Yam}
\affiliation{New England Complex Systems Institute, 
 24 Mt. Auburn St., Cambridge, MA 02138, USA}
\affiliation{Department of Molecular and Cellular Biology, 
Harvard University, Cambridge, MA 02138, USA}
\author{Mehran Kardar}
\affiliation{Department of Physics, Massachusetts Institute of Technology, 
 Cambridge, MA 02139, USA}

\begin{abstract}
We reconsider the persistence of information under the dynamics of 
the logistic map in order to discuss communication through a nonlinear 
channel where the sender can set the initial state of the system with
finite resolution, and the recipient measures it with the same accuracy. 
We separate out the
contributions of global phase space shrinkage and local phase space
contraction and expansion to the uncertainty in predicting and postdicting
the state of the system. Thus, we determine how the 
amplification parameter, the time lag, and 
the resolution influence the possibility for communication. 
A novel representation for real numbers is introduced 
that allows for a visualization of the flow of information between
scales. 
\end{abstract}

\maketitle
\section{Introduction}
When observing a dynamical system in a given state, 
one can ask several basic questions, two of which are, ``Where did it come
from?'' and ``Where is it going?'' These questions are especially
relevant when one wants to communicate a message by setting the initial
state of a system, whose state at a later time is then detected by the
recipient of the message. If the final state is completely uncertain, the
message is lost -- this is often the case in chaotic systems. On the other
hand, if all initial states converge to one final state, the recipient
cannot determine what message he was to receive -- this can occur in
dissipative systems. Many nonlinear systems have elements of both chaos and
dissipation; as an example, we choose the logistic map $f(x) = a x (1-x)$
\cite{Ott:Chaos}. We determine the relevance of state space shrinkage 
and expansion for all values of the amplification parameter $a$, and
discuss to what extent communication in the sense mentioned is possible.
Note that using a chaotic channel is not something 
the sender {\em chooses} to do -- he may
be forced to entrust the message to an unreliable medium. This
distinguishes our perspective  from 
other publications on communication through chaos such as in Refs.
\cite{Hayes:Communicating,Pecora:Synchronization,Bollt:Coding,
  Rulkov:Chaotic, Baptista:Cryptography}, 
where chaotic dynamics are used in order to amplify and
transmit small signals.

We will show that for very short time intervals between the initialization
and the measurement, the chaotic regime near $a= 4$
allows for optimal communication; for intermediate times, the bifurcation
points offer the best chances of deciphering the message; and for long
times, no information remains except in the bifurcation regime, where one
can distinguish between the branches of the cycle. Under all circumstances,
uncertainty about the time at which the 
system was initialized leads to additional losses of information. 

The degree to which an observation at one time determines the result of an
observation at a different time is given by the mutual information between
these observations, and the conditional information between them.
 The relevance of measuring information to characterize the
behavior of chaotic systems was realized more that 20 years ago
\cite{Shaw:Strange,Farmer:Information,Pompe:State,Pompe:Transinformation, Kruscha:Information} 
and discussed in various contexts 
\cite{Schittenkopf:Exploring,Deco:Information,Wiegerinck:Information}; 
some of the results
presented here have been either alluded to or derived before, especially in
Ref. \cite{Pompe:State}. However, for concrete calculations, most of these
references assumed that the system had already reached a stationary state
or attractor, whereas relaxation to the attractor plays a crucial role in
our analysis. Also, previous studies have focused on the case of
fully developed chaos at $a=4$; we discuss the behavior for all regimes of
$a$.

When inputs and outputs are measured with finite precision, it is useful to
have a representation of numbers that separates contributions on different
scales. We introduce a representation that may be superior to the usual
decimal or binary representations in this regard, 
and use it to visualize the flow of
information between scales.

Section \ref{SEC-info} provides an overview of 
information-theoretical concepts, and a derivation of equations needed 
to calculate the relevant quantities for generic maps. In
Sec. \ref{SEC-logmap}, we apply the formalism to the logistic map. 
The possibility of transmitting messages by initializing 
the system is discussed in Sec. \ref{SEC-commun}. In Sec. \ref{SEC-rep}, 
we introduce the {\em clockwork representation} of real numbers and use it to
illustrate the dynamics of the logistic map. 
Section \ref{SEC-summary} summarizes the results.

\section{Information-theory perspective}
\label{SEC-info}
\begin{figure}
  \epsfxsize= 0.65 \columnwidth
  \epsffile{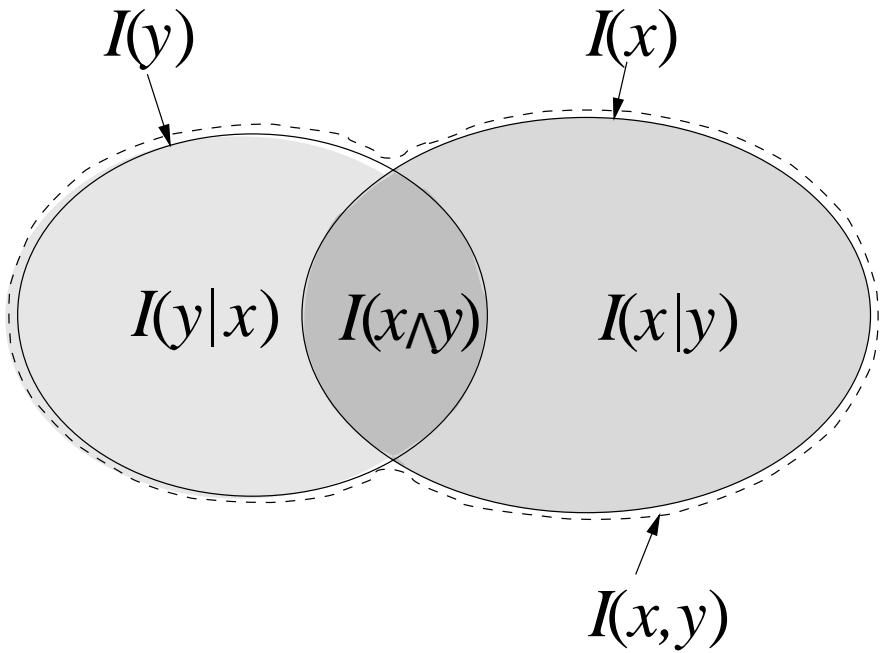}
   \caption{For two correlated events $x$ and $y$, the relations between
     the information $I(x)$, conditional information $I(x|y)$, joint
     information $I(x,y)$, and mutual information $I(x\wedge y)$ are
     illustrated in this diagram.}
  \label{LOG-infsketch}
\end{figure}

Since observations on physical systems can only be made with finite precision,
the outcome can be described with a finite number of digits. Each possible
distinct outcome (or elementary event) can be assigned a symbol 
that appears with a given probability, and
Shannon's definition of information \cite{Shannon:Mathematical} 
can be applied to these symbols,
resulting in a resolution-dependent, but finite, information. 
 We divide the
space of possible inputs and outputs $x\in [0,1]$ 
into $r$ non-overlapping bins $i$ of uniform width $1/r$, and assume
 that measurement precision is independent of
the value \cite{endnote18}.
%\footnote{This can be seen as defining what the observed quantity
%  is. Transformations commonly used to switch from one set of variables/
%  equations to another do not conserve this property.}
 Accordingly, we 
denote by $x_i$ the event ``a trajectory starts in bin $i$,'' and with
$y_j$ the event ``a trajectory ends in bin $j$.''
If $x_i$ is drawn from a probability distribution $P_x(x)$, the 
information associated with $x$ is 
\be
I(x) = -\sum_i P_x(x_i) \log_2 P_x(x_i);
\ee
the information $I(y)$ of $y$ can be calculated analogously from the
distribution $P_y(y)$. 
The conditional information, which is needed to specify the 
outcome $y$ given the input $x$, is
\be 
I(y|x) = - \sum_i P_x(x_i) \sum_j P_{y|x}(y_j|x_i) 
\log_2 P_{y|x}(y_j|x_j),
\label{LOG-condinf1}
\ee
where $P_{y|x} (y_j|x_i)$ denotes the conditional probability of $y_j$
occurring given that $x_i$ occurred.
All other quantities (mutual information $I(x\wedge y)$, 
joint information $I(x,y)$, and conditional information $I(x|y)$
) can be calculated from $I(x)$, $I(y)$, and $I(y|x)$,
 using the set-theoretical 
relations implied by
Fig. \ref{LOG-infsketch}. In particular, we make use of $I(y)=I(y\wedge
x)+I(y|x)$: to specify $y$, we need the mutual information $I(y\wedge x)$,
which represents information about $y$ that can be inferred
from knowledge of $x$, and the conditional information $I(y|x)$.

A {\em bijective} mapping between input and output ($x$ completely determines $y$
and vice versa) is achieved if $I(x\wedge y)= I(y) = I(x)$.
Note that $I(x\wedge y)\leq \max(I_x,I_y)$ -- if the space of possible states
shrinks when applying $y=f(x)$, there can be at most an {\em injective}
mapping ($x$ completely determines $y$, but not vice versa).

We first discuss $I(y)$, then the conditional information
$I(y|x)$, and consider what processes influence them. We then 
develop a formalism to derive analytic results for 
functions that are smooth on scales comparable to the resolution, 
e.g., if $y$ is the result of
applying the logistic map $f(x)=ax(1-x)$ iteratively for a small number of
times. The results for long times are discussed later
using a different approach.

\subsection{Total information: Global phase space considerations}
\label{SEC-phase}
Under any mapping, an ensemble of input values 
drawn from a given probability distribution
is generally mapped onto an ensemble of outputs that is described by another 
distribution. It is necessary to check whether one or the other requires
more information to describe an associated event (a distinct value of input or
output).

Sampling a continuous random variable many times with resolution $1/r$ is
equivalent to generating a histogram with bins of width $r$.
It is useful to separate the information needed to select an element  
from this histogram
into two contributions: one from treating the 
function underlying the probability distribution as continuous in $x$, 
and another from the act of separating the input space into $r$
bins. Let us say  that we have a probability density 
$p_x(x)$ living on $x\in [0,1]$. The information of this distribution, 
according to the usual 
definition \cite{Shannon:Mathematical}, is
\be
I_c = -\int dx\; p_x(x)\; \log_2 p_x(x). \label{LOG-infodef1}
\ee
The  discrete probability distribution of the histogram is 
$P_{x} (x_i) \approx p_x(i/r)/r$. The information
of events drawn from this discrete distribution is 
\be
I_d = -\sum_i P_x(x_i) \log_2 P_x(x_i)
 \approx -\int dx\; r\;( p_x(x)/r) \log_2(p_x/r)=  I_c + \log_2 r.  
 \label{LOG-infodef2}
\ee
Replacing the sum by the integral is valid as long as $p_x(x)$ is reasonably
smooth over the range of one bin (which is only roughly valid for the
distributions that we discuss in Sec. \ref{SEC-short}). 

If the input is drawn from a known probability distribution $p_x(x)$, 
the output $y$ follows a probability distribution $p_y(y)$ that can be
calculated using the rules for transforming probability
distributions \cite{VanKampen},
\be
|p_x(x) dx| = |p_y(y(x))dy|, \Rightarrow p_y(y) = \sum_{\alpha} p(x_{\alpha}(y))
\left |\frac{dx}{dy} \right |,
\label{LOG-probmap} 
\ee
the sum being over all $x$ which map onto $y$. Under all one-dimensional
 chaotic maps, including the logistic map, two or more input values are
mapped onto the same output, a property known as {\em folding}.

\subsection{Conditional information: Local expansion and shrinkage 
of phase space}
\label{SEC-expand}
We now study the conditional information $I(y|x)$. 
While the last section dealt with global properties of the map,
here we are averaging over a local property -- 
given some $x_i$, we ask, ``how much can we know about the
output?'', which is independent of the behavior of the map for other input 
values $x_{k}$.

\begin{figure}[h]
  \epsfxsize= 0.65 \columnwidth
  \epsffile{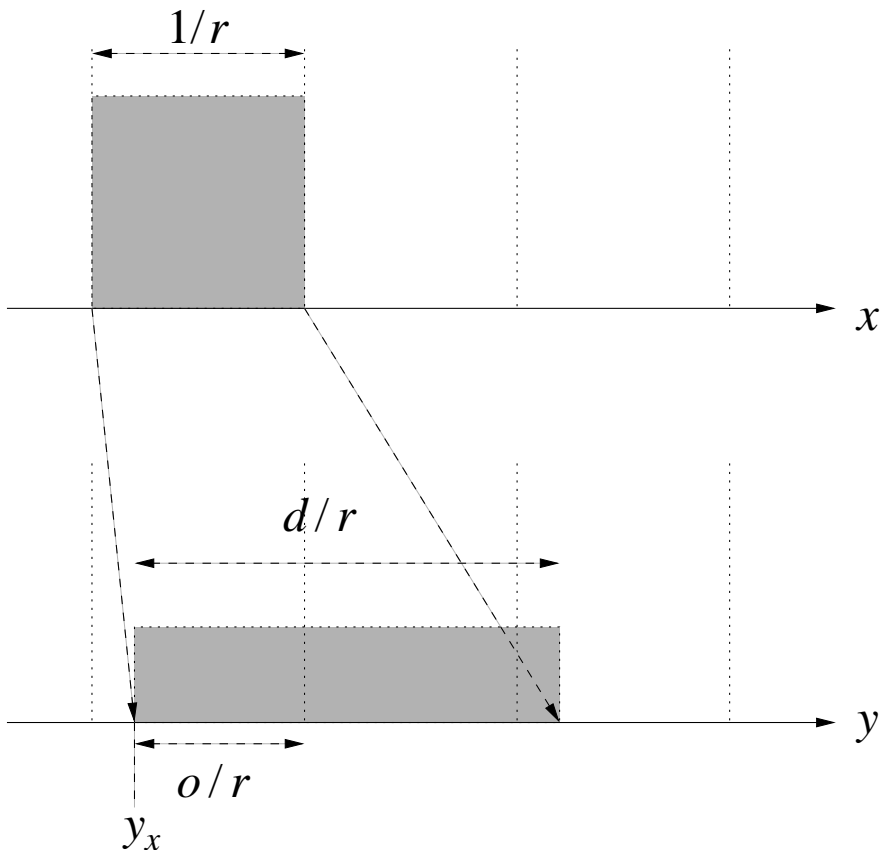}
   \caption{Assuming small bins, trajectories from within one bin in $x$ are
     mapped uniformly onto one or more bins in $y$.}
  \label{LOG-mapping}
\end{figure}

Let us denote the local conditional information as
\be 
I_l(y|x_i) =\sum_{j} P_{y|x}(y_j | x_i)  \log_2 P_{y|x} (y_j |x_i).
\label{LOG-Ilocal}
\ee
As long as bins are small compared to changes in $dy/dx$,
trajectories starting from  points in the interval  
$[x_i, x_i + 1/r]$
are uniformly distributed over an interval $[y_x, y_x+d/r]$, where 
$d = |dy/dx|$. The uncertainty about the outcome is determined by the
overlap of this interval with the bins (as sketched in Fig. 
(\ref{LOG-mapping})). 
To account for this, we average
over the offset $o\in [0,1]$ which specifies where $y_x$ is within a bin.

If a percentage $o$ of a bin is covered with trajectories, the contribution
of that bin to the sum in Eq. (\ref{LOG-Ilocal}) is given by 
$I_p(o) = -(o/d) \log_2 (o/d)$. Let us look first
at the case $0<d<1$. In that case, either the covered interval is entirely
within one bin (if $o>d$) -- then the conditional information is $0$ -- or
the trajectories are split between two bins, resulting in a non-vanishing
conditional information. Averaging over $o$ gives
\be
\langle I_l(y|x)\rangle_o =  \int_0^d\left (I_p(o)+I_p(d-o)\right)\, do + \int_d^1 0\; do
=\frac{d}{2 \ln 2}. \label{LOG-stretch1}
\ee
If $1\leq d < 2$, trajectories can be spread out over two or three bins,
depending on the offset:
\be
I(y|x) = \left\{ \begin{array}{ll} 
      I_p(o) + I_p(d-o)  &\mbox{~~~for~~~}o>d-1                    \\
      I_p(o) + I_p(1) + I_p(d-1-o)  &\mbox{~~~for~~~}o\leq d-1         
  \end{array} \right. .
\ee
Averaging yields
\be
\langle I_l(y|x)\rangle_o = \log_2 d + \frac{1}{2 d \ln 2}.  \label{LOG-stretch2}
\ee
Equation (\ref{LOG-stretch2}) thus has two contributions: 
a term logarithmic in $d$
to account for the bins that are fully covered, and a term from the two
partially covered bins, whose impact decays as $1/d$. One can show that 
Eq. (\ref{LOG-stretch2})  is valid for any value of $d>1$.

To find the average conditional information $I(y|x)$, we can now
sum  over $x$, with $d(x)=|dy/dx|$: 
\be
I(y|x) \approx \int \langle I_l(y|x) \rangle_o dx.
\label{LOG-confinfint}
\ee
 
%Note that the derivation assumed that the offset $o$ was randomly
%distributed. It can be expected that this works well for large numbers of
%input bins with functions with a non-vanishing curvature $d^2y/dx^2$ 
%such as the logistic map), but gives quantitatively wrong results for 
%piecewise linear maps such as $y = 2 x \bmod 1$.

\subsection{Folding}
It is well known that chaotic iterative maps require a folding mechanism to
compensate for stretching of phase space. E.g., in the case of the logistic
map, the two branches of the parabola map two input points onto the same
output. Clearly, through this process, information about the original
state is lost.
In the framework presented so far, this is not accounted for
explicitly; however, it is contained implicitly in the conditional
information. For example, when comparing the identity function $f(x)=x$ with
the shift map $f(x)= 2x \bmod 1$ (which is chaotic and has folding), 
both map the unit interval
uniformly onto itself and thus have the same global information $I(y)$.
However, the latter has a larger average local slope and thus, according to
Eq. (\ref{LOG-stretch2}), a larger conditional information, leading to a
smaller mutual information between input and output. In this case, the
uncertainty  in prediction generated through stretching  is the
same as that in postdiction through folding: going forward in time, one
does not know in which of two adjacent bins the output will be, whereas
looking back, one has two possible input bins that are well separated, 
one with $x<1/2$ and one with $x>1/2$.

\section{Application to the Logistic Map}
\label{SEC-logmap}
\subsection{Basics of the Logistic Map}
We briefly review
fundamental properties of the logistic map $f(x)=a x (1-x)$ when 
used as an iteration (i.e. $x_{t+1}= f(x_t)$). For $a\leq 1$, there is one
(stable) fixed point, namely $x_t = 0$. Between $1<a<3$, the only stable
fixed point is $x_t = 1 - 1/a$. At $a=3$, this becomes unstable and gives way
to a stable 2-cycle. What follows is a succession of bifurcations
($n$-cycles split into $2n$-cycles) at $a\approx 3.4493$, $3.54396$,
$3.56438$ etc., until the cycles merge into a continuous chaotic attractor
at $a_c \approx 3.569946$ (see also the bifurcation diagram at the top of
Fig. \ref{FIG-convfour}). The chaotic regime is interrupted by smaller and
larger windows of periodic behavior. At $a=4$ one gets what is 
often referred to as ``fully developed chaos;'' there, 
the chaotic attractor spans the interval $[0,1]$.

The Lyapunov exponent, which determines whether two trajectories starting
from nearby initial conditions converge or separate exponentially 
\cite{Eckmann:Ergodic}, is
negative in the fixed point and bifurcation regime, becomes zero at the
bifurcation points, and is positive in the chaotic regime. 

We now apply the formalism developed in Secs. \ref{SEC-phase} and
\ref{SEC-expand} to the first few iterates of the logistic map
$f(x)= a x(1-x)$, then explain the long-time behavior, and give numerical
 results on intermediate times.

\subsection{Short time behavior}
\label{SEC-short}

Let us consider the probability distribution $p_y(y)$ of the first
iterate, starting from a uniform distribution. There are two 
symmetric branches of $y(x)$, and Eq. (\ref{LOG-probmap}) gives
\be
p_y(y) = \left \{\begin{array}{cl}  
      \frac{2}{\sqrt{a}}\frac{1}{\sqrt{a-4y}} &\mbox{~~~~for~~~}
      y<\frac{a}{4} \\
      0 & \mbox{~~~~for~~~}y\geq \frac{a}{4}      
    \end{array}\right. .
\ee
The information of this distribution can be evaluated using Eqs. 
(\ref{LOG-infodef1},\ref{LOG-infodef2}):
\bea
I_c(y) &=&  - 1 -  1/\ln 2 + \log_2 a < 0, \\ \nonumber
I(y)&=& -1 -1/\ln 2 + \log_2 a + \log_2 r. \label{LOG-yinfo}
\eea
In contrast, the information of the uniform 
distribution was $I_c(x)=0$ and $I(x)=\log_2(r)$, which means that information 
about the state of the system was lost --  $1/\ln 2 -1 \approx
0.44$ bits for $a=4$, and more for $a<4$. 

\begin{figure}
  \epsfxsize= 0.65 \columnwidth
  \epsffile{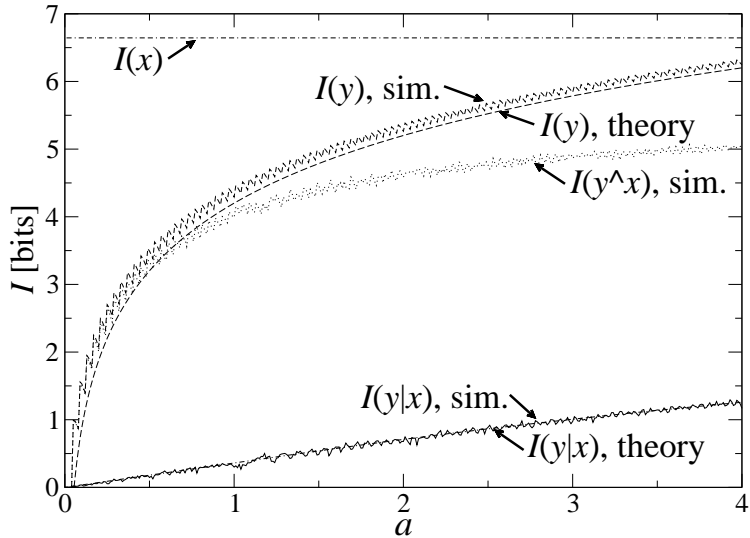}
   \caption{Conditional, mutual and total information 
     of input and output for $r=100$ for one
     time step, compared to Eqs. (\ref{LOG-yinfo}) and (\ref{LOG-condinfo}).}
  \label{FIG-infoT1}
\end{figure}

The conditional information can be derived from integrating 
Eqs. (\ref{LOG-stretch1}) and (\ref{LOG-stretch2}) over the input space.
In the first iterate of the logistic map, the
derivative is smaller than $1$ for $x\in [(1-1/a)/2, (1+1/a)/2]$, and
larger in the rest of the domain. Choosing boundaries appropriately and
making use of the symmetry of the system, we obtain
\be
\int_0^1 P(x) I_l(y|x) dx = \frac{a}{4\ln 2}
\ee
for $a<1$, and 
\bea
\int_0^1 P(x) I_l(y|x) dx&=& \frac{1}{a \ln 2}\left[ \int_1^a \left(\ln x
  +\frac{1}{2x}\right )dx  + \int_0^1 \frac{x}{2} dx\right] \nonumber\\
&=& \frac{1}{a \ln 2} \left [(a+ 1/2) \ln a + \frac{5}{4} -a \right]
\label{LOG-condinfo}
 \eea
for $a\geq 1$. This agrees well with numerical results for finite
resolution, as seen in Fig.  
\ref{FIG-infoT1}. It should be pointed out that $I(y|x)$ does not
explicitly depend on the resolution, in contrast to $I(y)$.

In Fig. \ref{FIG-infoT1}, $I(y|x)$ represents the uncertainty generated by
stretching and compressing; $I(x)-I(y)$ represents the uncertainty through
shrinking of phase space; and $I(x)-I(y\wedge x)$ is the average
information necessary to reconstruct $x$ from knowledge of $y$.
The mutual information $I(y\wedge x)$ is the amount of information about
the initial state retained after the mapping.

\begin{figure}
\begin{minipage}{0.49 \columnwidth}
  \epsfxsize= 0.98 \columnwidth
  \epsffile{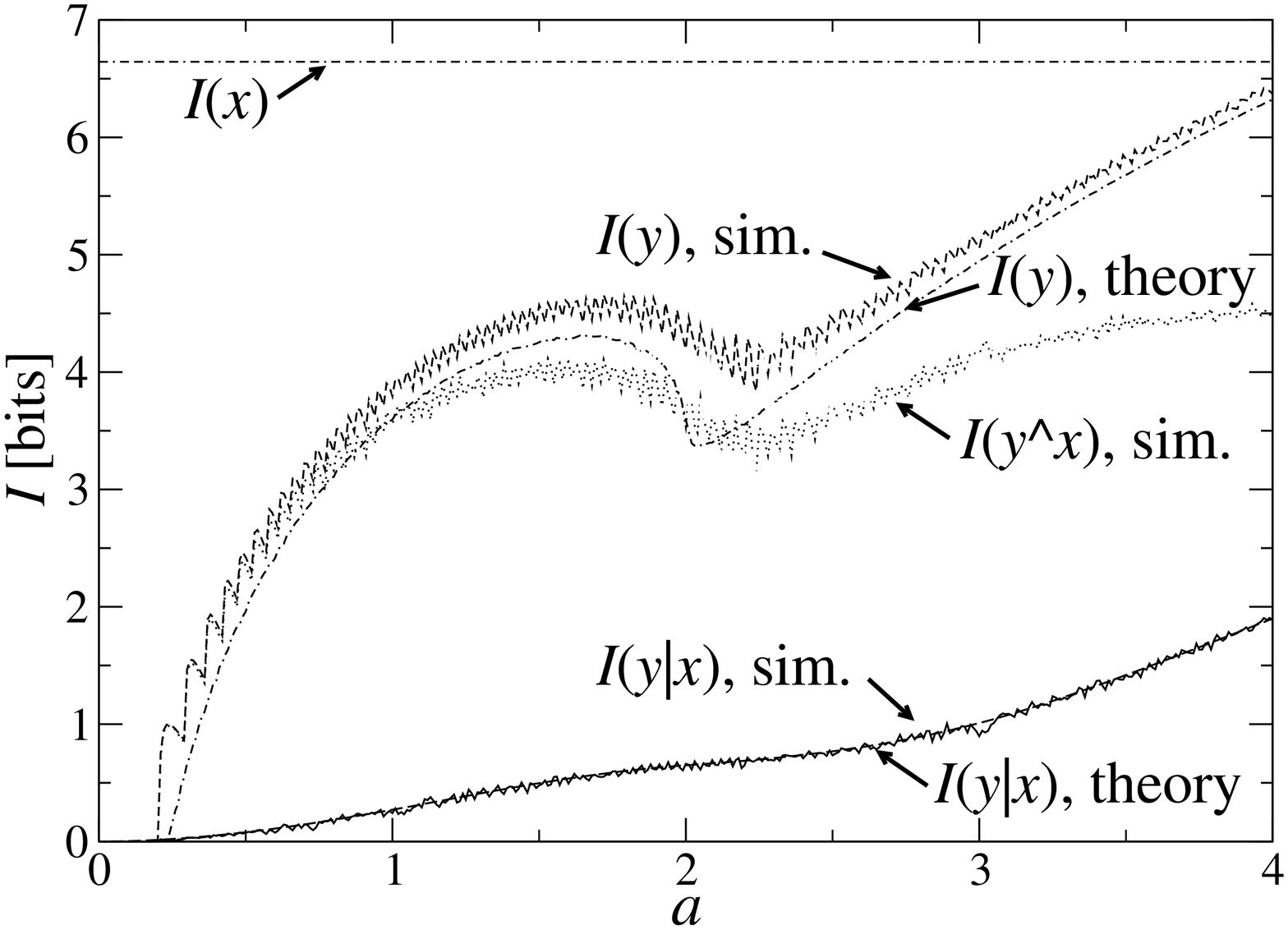}
 \end{minipage}
\begin{minipage}{0.49 \columnwidth}
  \epsfxsize= 0.98 \columnwidth
  \epsffile{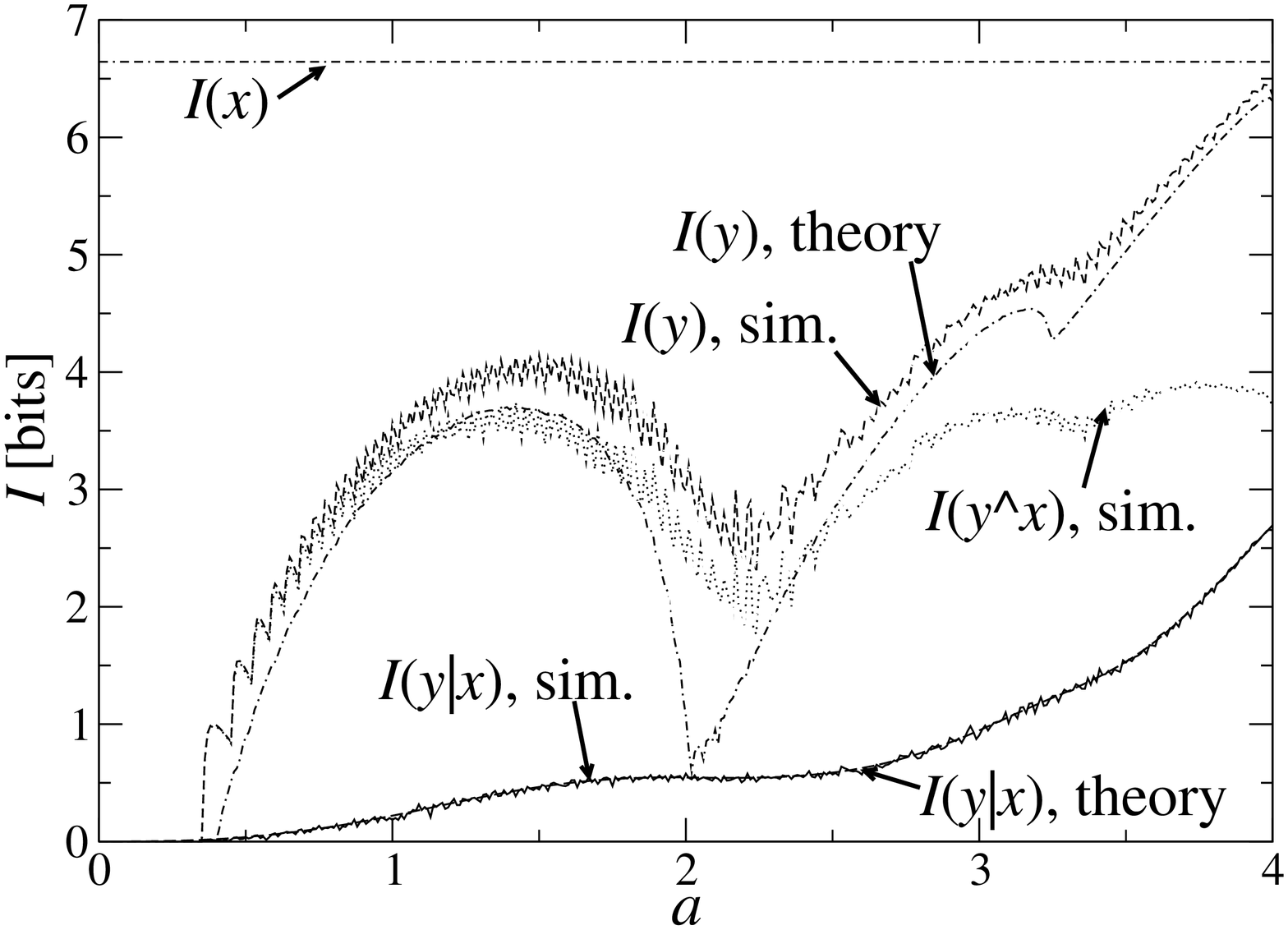}
\end{minipage}
   \caption{Conditional, mutual and total information 
     of input and output for $r=100$ for two and three
     time steps, compared to results of numerical integration. $I(y|x)$
     does not depend strongly on $r$, whereas the other quantities include an
     additive term of $\log_2 r$.}
  \label{FIG-infoT23}
\end{figure}

The mutual information can be calculated by numerical integration over 
Eq. (\ref{LOG-stretch2}) for the second and third iteration of the logistic
map, and good agreement with simulations is found, as shown in Fig. 
\ref{FIG-infoT23}. For higher iterates, 
numerical integration becomes difficult.
Numerical integration over the probability distribution of outputs also
becomes less accurate, and the approximation made in
Eq. (\ref{LOG-infodef2}) becomes visibly wrong for resolutions as coarse 
as $r=100$. 

While Figs. \ref{FIG-infoT1} and \ref{FIG-infoT23} do not show a clear
distinction between the fixed point/cyclic regime and the chaotic regime,
one can see that conditional information (i.e., uncertainty
generated by the dynamics) increases with $a$, whereas $I(y)$ develops
dips. For example, the dip at $a\approx2$ represents rapid convergence to the
fixed point far from the bifurcation points $a=1$ and $a=3$.
Correspondingly, $I(y\wedge x)$ is no longer monotonic in $a$ --
several maxima of conserved information emerge.

\subsection{Long-time behavior} 
\label{SEC-longtime}
\begin{figure}
  \epsfxsize= 0.65 \columnwidth
  \epsffile{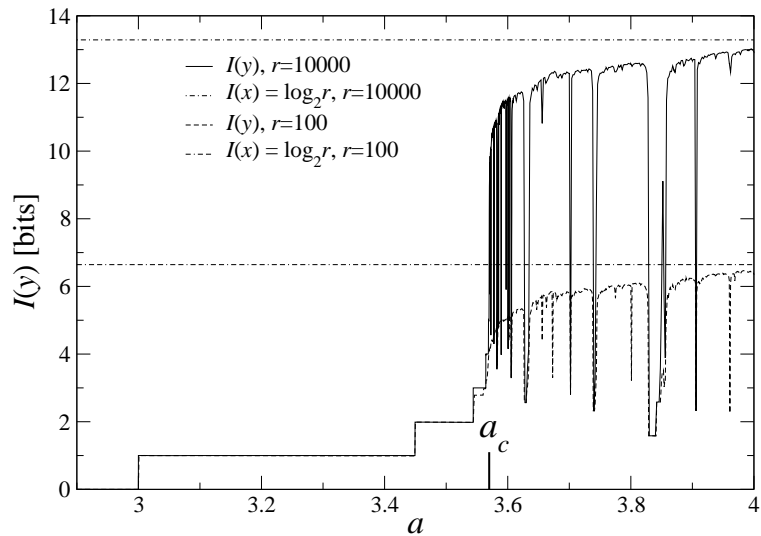}
   \caption{Information $I(y)$ of the output distribution for $r=100$ and 
     $r=10000$,
     for $t\rightarrow \infty$, compared to the input information $I(x)$.}
  \label{FIG-yinfo}
\end{figure}

For very long times, some simple statements about the information in the
output can be made: if the map has a single fixed point ($0<a<3$), $I(y)=0$. 
For a cycle of length $2^n$, $y(t)$ can be defined completely 
by stating what branch of the cycle it is on; the information 
is therefore  $n$ bits if the resolution is fine enough to resolve each
branch of the cycle, and each branch has an equally large basin of
attraction, and smaller otherwise. In the chaotic regime, 
there is a probability distribution filling a finite fraction of the
interval $[0,1]$ for most values of $a$, and cycles of various lengths in
certain periodic windows.
The continuous information $I_c$ will therefore be less than one, 
and the information at finite resolution less than or equal to $\log_2 r$, with visible
dips in the periodic windows. Numerical
results of $I(y)$ for $r=100$ and $r=10000$ shown in Fig. \ref{FIG-yinfo} 
demonstrate these features.

\begin{figure}
  \epsfxsize= 0.65 \columnwidth
  \epsffile{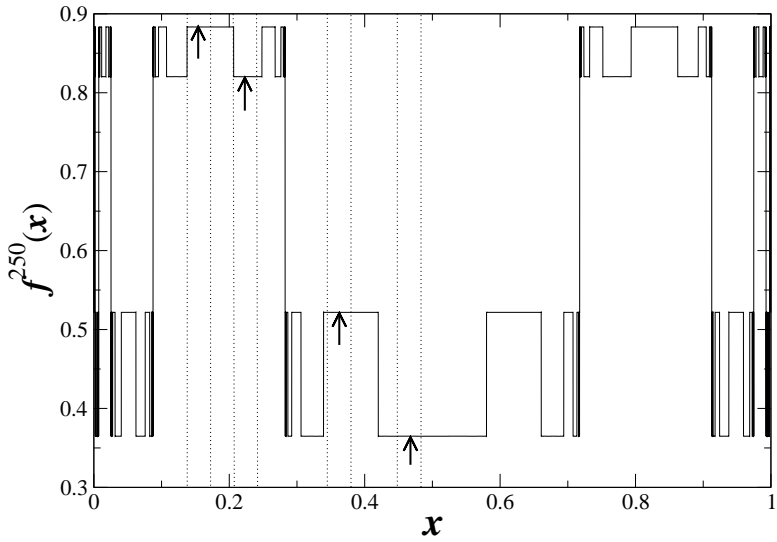}
   \caption{The 250th iterate of the logistic map at $a=3.54$, as an
     example of the long-time dynamics.  
     The
     dominant values are the branches of the 4-cycle; one sees the fractal
   structure of the basins of attraction of each branch.  
   Iterates at times $250+4n$ look indistinguishable. The dotted lines
   indicate input bins at a resolution of $r=29$ that map uniquely onto one
 branch of  the cycle (see Sec. \ref{SEC-commun}).}
  \label{FIG-basins}
\end{figure}

The mutual information is at most as large as the total information, therefore it
will be $0$ as well for $a<3$. In the bifurcation regime, dynamics are
fairly predictable. The basins of attraction for each branch of the cycle
are fractals, reminiscent of Cantor sets, as shown in Fig. \ref{FIG-basins}. 
If the bins are small enough such that most bins map exclusively to one branch of a $2^n$-cycle, 
the mutual information is of order $n$ bits. 

In the chaotic regime, information about the original state is lost at a
rate approximately equal to the Lyapunov exponent 
\cite{Shaw:Strange,Pompe:State}, which here is between 0 and 1; we
therefore expect mutual information to be $0$ after 
$\mathcal{O}(\log r)$ time steps. Note that this affects prediction as well
as postdiction: even though $I(y)$ is not much smaller than $I(x)$ for
$a=4$, in the absence of mutual information, it is as impossible to tell
where the system came from as where is is going.

\subsection{Intermediate times}
\begin{figure}
  \epsfxsize= 0.65 \columnwidth
  \epsffile{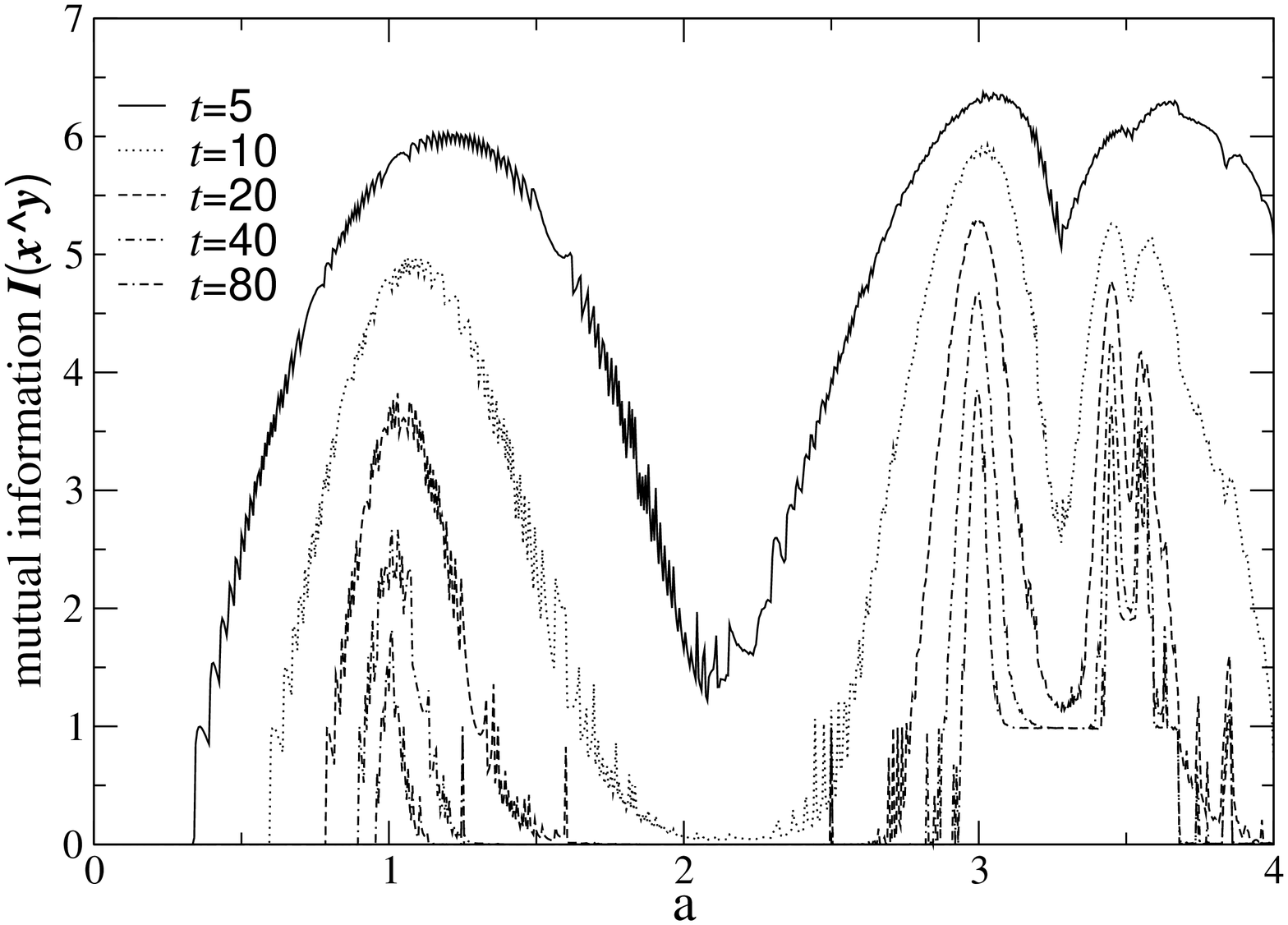}
   \caption{Mutual information between input and output for different
     numbers of iterations $t$, for a resolution $r=1000$. For smaller
     resolutions, curves look similar, but more jagged. Also, the maximum
     of the curves has an additive term of $\log_2 r$.}
  \label{FIG-inttime}
\end{figure}

Figure \ref{FIG-inttime}
shows the mutual information for $r=1000$ at various intermediate 
times $t$, measured by
scanning input space with a step width small compared to the bin width.
Apart from the long-time features explained in Sec. \ref{SEC-longtime}, one
notices several peaks. The narrow peaks (e.g. near $a=2.5$) 
occur when the fixed point is very close to the boundary between two
bins, such that small deviations from the fixed point lead to ambiguities
in the outcome. They change position if the binning is chosen differently.

The wider peaks at $a=1$, $3$, and $3.54$ correspond to the bifurcation
points. There, the Lyapunov exponent is 0;  deviations from the fixed point or
cycle decay like a power law rather than exponentially.
In the chaotic regime, the mutual information quickly drops to $0$, apart
from peaks that correspond to periodic windows (e.g. at $a\approx3.83$).

\section{Communication through a logistic map channel}
\label{SEC-commun}
We now interpret and expand 
the results of previous sections with a view to the problem of communication 
in the following scenario: a sender A wants to transmit a message to a
receiver B by setting the initial state of the logistic map with some finite
resolution. B receives the state after $t$ iterations of the logistic map
and interprets it. For what values of
$a$ and $t$ can A expect any degree of transmission, and what resolution do
A and B need?

The scenario may seem contrived; however, setting the initial state of some
physical system (like a sheet of paper, or a hard disk) in the hope that
someone will be able to read it {\em is} the usual way of transmitting
messages over long times. Usually people choose systems whose dynamics are
somewhat stable to perturbations and slow compared to the time $t$, but they
may not always have that choice. 
Let us look at two different communication problems.

\subsection{Reconstructing the initial state}
 In this scenario, 
A wants B to know the state that A started the system
in. The problem is then essentially one of postdiction for the recipient,
and the relevant quantity is the difference between the mutual information
$I(y\wedge x)$
 and the input information $I(x)$. Surprisingly, for very short
times, Figs. \ref{FIG-infoT1} and \ref{FIG-infoT23} show that 
B has the best chances of postdicting the initial state in the
chaotic regime near $a=4$ -- the loss of information through chaos is
not as significant as that through phase space shrinking in the low-$a$-regime.
Note that at least one bit is lost: since $f(x)$ is symmetric around $x=1/2$, it is impossible to tell whether the system was started on the left or right branch.

For intermediate times, at high $a$, chaotic dynamics eliminates all 
information about the initial state; so does fast convergence to a single fixed
point for small $a$. As Fig. \ref{FIG-inttime} shows, 
B's situation is best if $a$ is close to one of the
bifurcation points, where convergence follows a power law rather than an
exponential.

In the limit of very long times, when the system has converged to its
attractor, the only regime where information about the initial state
persists is the bifurcation regime. What value of $a$ gives optimal
transmission depends on the resolution: the recipient has to be able to
resolve all branches of the cycle to make full use of the remaining 
information.

In all three time regimes, it is important for the recipient to know the
precise time at which the system was initialized. The information required
to specify the time lag has to be included in the conditional information 
$I(x|y)$. To give a simple example: in the long-time regime, if the time
is either $t$ or $t+1$ with probability $1/2$, an additional bit of
information is required to reconstruct the initial state.

\subsection{Determining the final state with certainty}
The second communication problem is this: A only wants to send B a
message that B can decode with certainty: A chooses the initial state
(again with resolution $r$) such that all trajectories from that state end
in one output bin. The relevant questions are now, what resolution do A and
B need to specify at least two different final states for various times 
and amplifications; and, given a certain resolution, how many distinct final
states exist that can be achieved with certainty by choosing an appropriate
initial state? This is a problem of prediction on the part of the sender.

The answer is clearest for long times: as described above, the only regime
where information persists and communication is possible at all
is that of cyclic behavior. The sender needs
to identify input bins that lie completely in the basin of attraction of
one branch of the cycle, such that each initial value in the bin leads to 
the same final value.  One such bin should be found 
for each distinct branch, and the resolution must be 
sufficient for the recipient to identify each branch (see
Fig. \ref{FIG-basins} for an example of such a set of bins). Numerics
show that the latter constraint is weaker than the first: it is the
sender's resolution that limits communication.
We find that a resolution of 7 bins is sufficient for the 2-cycle at values of
$a= 3.5$; thus, $\log_2 7 \approx 2.8$ bits are needed to specify the
input to transfer one bit to the recipient. For longer cycles, the ratio
can be more efficient: 29 bins are enough to resolve the 4-cycle at
$a=3.54$, yielding $4.86$ bits of input for 2 bits of output. At $a=3.562$,
a resolution of 122 specifies each branch of the 8-cycle, giving 3 bits of
output for $6.93$ bits of input. (We assume that $a$ is constant and
precisely known to the sender and recipient.)

\begin{figure}
  \epsfxsize= 0.65 \columnwidth
  \epsffile{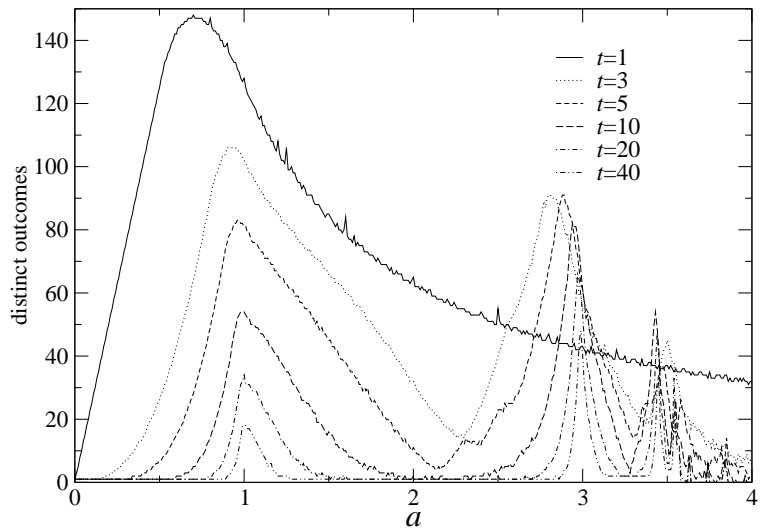}
   \caption{The number of distinct final states that can be reached with
     certainty by choosing an appropriate initial state. Both sender and
     recipient use $r=1000$.}
  \label{FIG-finalstates}
\end{figure}

In the short and intermediate time regimes, the number of distinct final
states that can be reached with certainty increases roughly linearly with
resolution. The slope is a function of both $t$ and $a$, 
and its value indicates the amount of information
lost in the channel. Fig. \ref{FIG-finalstates} shows the number
of distinct states at $r=1000$ for different times and amplifications.
Some features are similar to Fig. \ref{FIG-inttime}: for longer times,
peaks at the bifurcation points emerge, representing the slow loss of
information. Between the bifurcation points, plateaus at
values of 2 and 4 can be seen for $t=40$. In the chaotic regime,
the number of predictable final states goes to zero with increasing time.

The most surprising feature of Fig. \ref{FIG-finalstates} is that 
the curve for $t=1$ is not monotonic, and even drops below those for
longer times. The reason is that the first iterate has a slope greater than
$1$ for
most $x$ values at high $a$, which makes unique prediction impossible.
Further iterations can re-compress parts of the state space that were 
stretched in the first iteration, leading to a better match between input
and output bins.

\section{Scale-resolving representations of real numbers}
\label{SEC-rep}
Dissipation and chaos have a common aspect: in both cases the dynamics 
makes a connection between large and small scales. Chaotic dynamics amplify
small differences in the initial states until they reach macroscopic
proportions, whereas dissipative dynamics shrink differences until they
vanish below the threshold of perception.
To represent this adequately, we 
first have to make clear what we mean by
information on different scales.

Let us consider real numbers $x$. On one hand, each real number can be
represented by one point on the real axis  -- it is a
one-dimensional quantity. On the other hand,  in the usual descriptions
(decimal, binary etc.), real numbers are represented by a set of integers
that stand for different scales -- the scales of 1s, 10s, 100s, etc. 
This makes sense because it reflects what happens when some $\Delta x$ is
added to $x$: the digits of $x$ are strongly affected for all scales finer
than the scale of  $\Delta x$, and weakly 
affected for coarser scales.

There are two problems with the usual representations: first, they are 
{\em discontinuous}: a small change on a fine scale will have no effect at
all on coarser scales most of the time, but a dramatic effect in rare cases
(such as when $0.001$ is added to $0.999$). This is a necessary side effect
of using discrete (integer) representations on each scale.
Second, they do not lend themselves to simple modification under
{\em multiplication}: multiplication is basically a convolution of the
representations of the two factors. 
Whereas multiplications with numbers that have a 
simple representation in the chosen base gives a shift
(e.g. multiplication with 10 in the decimal representation just shifts the
decimal point), all other factors lead to changes throughout the scales.
It may therefore be worthwhile to think about alternative representations
for which the information content of various scales is easy to visualize.  

First, we need to decide what properties we expect the representation to
have. At a very basic level, each real number should map onto one
representation, and each representation should map onto at most one 
real number. (Since we want to include the option of representing one real
number by a set of real or complex numbers, 
a bijective mapping is not generally
possible.) Also, elements of the representation 
corresponding to finer scales should not
include the information at coarser scales -- otherwise they would not be
specific to their scale. One way of achieving this is using periodic
functions, with the period equal to the scale to be studied.

\subsection{The Clockwork Representation}
\begin{figure}
  \epsfxsize= 0.65 \columnwidth
  \epsffile{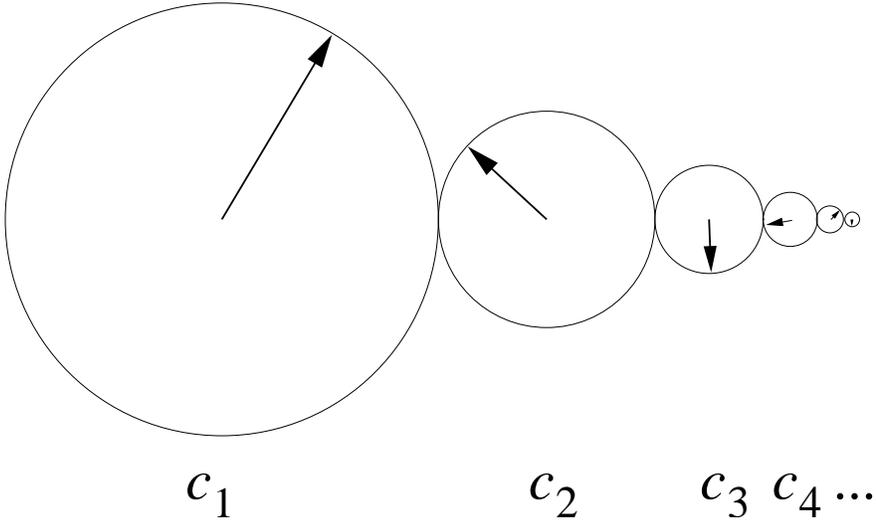}
   \caption{The clockwork representation; each digit represents one cog.}
  \label{clockworkfig1}
\end{figure}

Using the most natural periodic function, one obtains what we call 
the {\em clockwork representation} (CR).
It maps each real number $x$ onto  set of complex numbers 
\be
c_j(x) = \exp(2\pi \im\; 2^j x),
\ee
with $j$ identifying the scale. The base $2$ is was chosen in analogy to
the familiar binary representation  -- any number
other than 1 can also be used. The term ``clockwork representation'' was chosen because
each digit can be thought of as a cog in a clockwork of consecutively
smaller cogs, as depicted in Fig. \ref{clockworkfig1}
 -- two turns of $c_2$ cause one turn of $c_1$, but four turns
of $c_3$. Using a different base is equivalent to using cogs of a different
radius ratio.

We can use the usual laws for exponential functions to see how addition and
multiplication of numbers carries over to their clockwork representation:
\bea
c_j(x+y) &=& \exp(2 \pi \im\; 2^j x) \exp(2 \pi \im\; 2^j y) = 
c_j(x)\,c_j(y);\\
c_j(ax) &=& \exp(2 \pi \im \; 2^{j + \log_2 a} x) = c_{j+\log_2 a} x.
\eea
The second line appears a little problematic for two reasons: first, 
it introduces
an asymmetry between the object under consideration $x$ and the factor $a$
-- the result might as well have been written as $c_{j+\log_2 x} a$.
This is, in some cases, desired: there is often a conceptual difference
between the dynamical variables and the parameters of a model.
Second, scales  become continuous rather than discrete. However,
this is more of an advantage rather than a disadvantage. 
The outcome is well-defined even for non-integer scales, 
as opposed to the case of discrete
representations: one can write a number in base 2 or base 3, but not base 
$2.5$. 

The CR gives a set of complex numbers; while it is clear how to do
mathematical operations on them, it is not completely obvious how to 
{\em display} them. It can be argued that usually one is interested in the
imaginary part: it gives $0$ for scales much coarser than that of the number
under consideration, and it gives $0$ for fine scales if $x$ is a power of
$2$, much like the bits in a binary representation would.

\begin{figure}
  \epsfxsize= 0.65 \columnwidth
  \epsffile{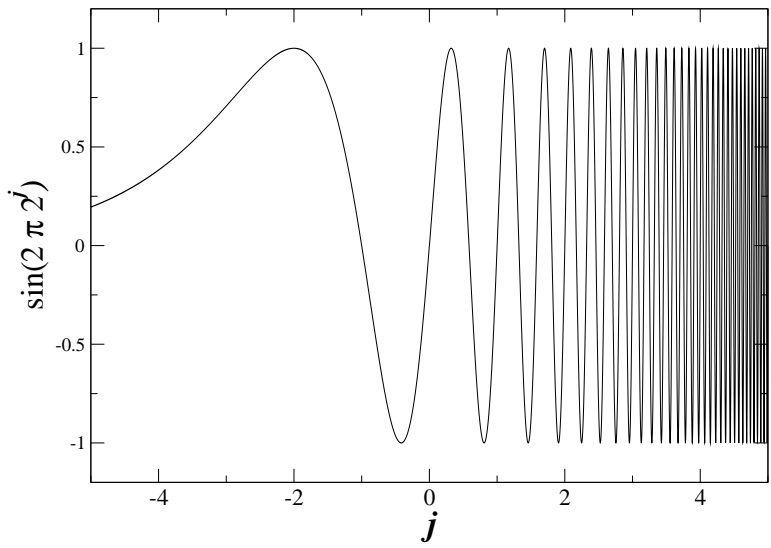}
   \caption{The clockwork representation of the number 1: 
     each digit corresponds to the
     value of this function at integer values of $j$.}
  \label{clockworkfig2}
\end{figure}

Another way to think about the CR is to look at the continuous 
function that generates the digits, as in 
Fig. \ref{clockworkfig2}: it shows $\sin(2 \pi 2^j)$, which is the
imaginary part of the CR of the number $1$. It has an exponential tail
toward coarse scales and an oscillating part with a frequency that
increases exponentially toward fine scales. The CR picks out the values of
this function at integer values of $j$. For any other number $x$, shift the
curve  to the left by $\log_2 x$, and again pick the values at integer
values of $j$.

\subsection{Using the CR in the logistic map}
\begin{figure}
  \epsfxsize= 0.65 \columnwidth
  \epsffile{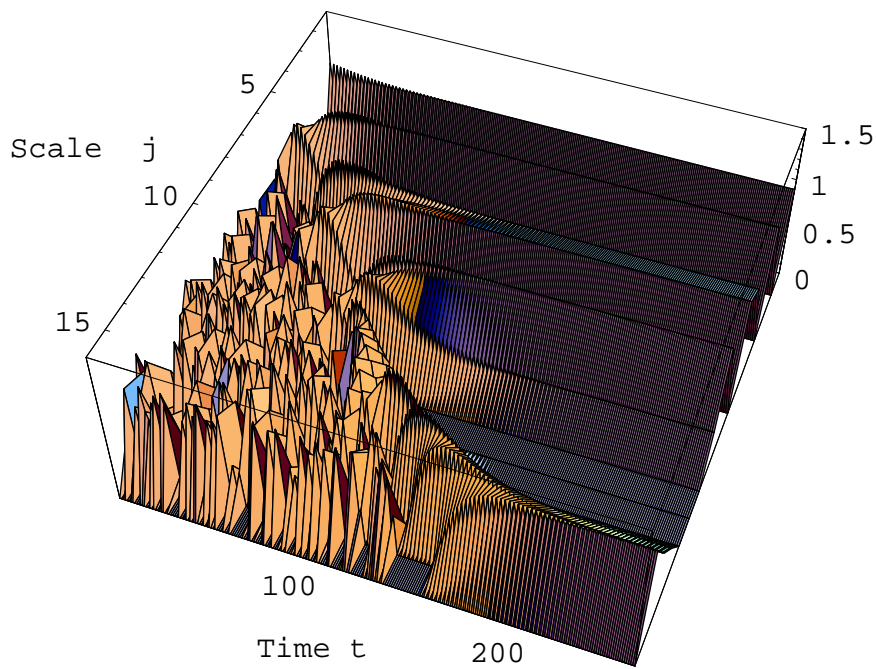}
   \caption{Imaginary part of the CR  of the $t$th iterate of the
     logistic map, at $a=2.95$. Deviations from the fixed point travel
     ``downstream,'' to finer scales.}
  \label{FIG-conv}
\end{figure}

\begin{figure}
  \epsfxsize= 0.8 \columnwidth
  \epsffile{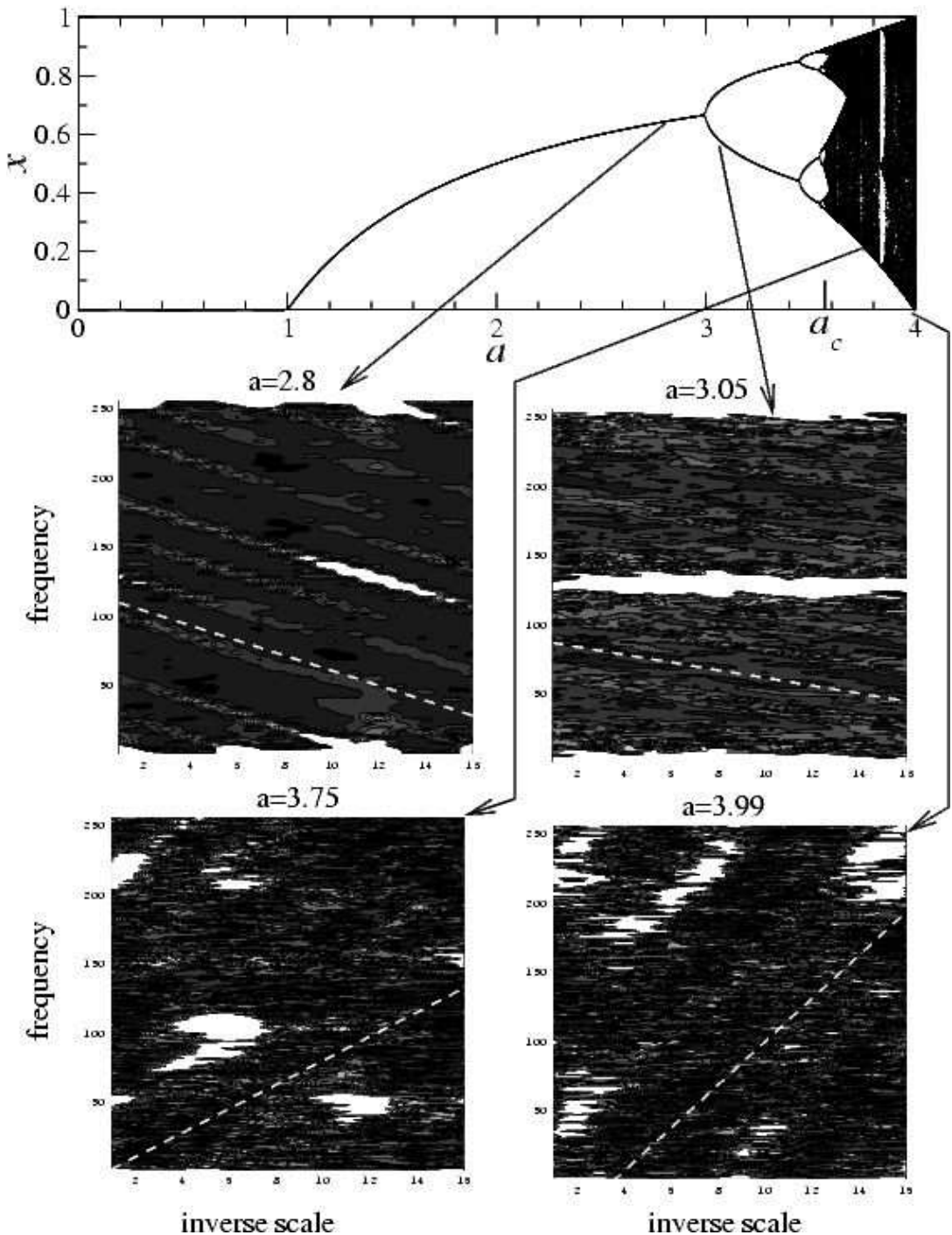}
   \caption{Squared amplitude of the Fourier transform of the CR, averaged
     over more than $100$ initial conditions, at $a=2.8$, $3.05$, $3.7$, and
     $4.0$. Dashed lines indicate a slope $\lambda/\ln 2$, where $\lambda$ 
     is the Lyapunov exponent.}
  \label{FIG-convfour}
\end{figure}

As explained in the previous sections, the clockwork representation can
give an impression about how numbers change on different scales of
resolution. This can be applied to illustrate the behavior of the logistic
map. The most obvious case is that of convergence to a fixed point or
cycle, as shown in Fig. \ref{FIG-conv}: for $a=2.95$ (slightly below the
first period doubling), a random initial state (left side) converges to 
the fixed point (right side); differences from the final state occur on
finer and finer scales as time progresses, resulting in a ridge traveling
to finer scales. 
This ridge also shows up in a two-dimensional Fourier transform of the CR,
seen in Figs. \ref{FIG-convfour}: 
for $a=2.8$, the absolute squared Fourier
transform of the CR shows diagonal ridges. To make the effect clearer, the
plot averaged over 100 uniformly distributed initial conditions. For
$a=3.05$, one sees a horizontal stripe in the center of the figure 
in addition to the diagonal
structures, indicating that the final state has a period of 2.

The slope of the diagonal structures during convergence to a limit cycle 
is directly proportional to the Lyapunov exponent $\lambda$: during each time
step, the deviation from the limit cycle diminishes by a factor of
$\exp{\lambda}$. In a plot of $\log_2 x$ vs. $t$ such as Fig.\ref{FIG-conv}, 
the ridge therefore has a slope of $\lambda/\ln 2$. In a Fourier transform
plotted as frequency vs. inverse scale, the ridge causes diagonal
structures of the same slope. To illustrate this, dashed lines of slope
$\lambda/\ln 2$ are shown in Fig. \ref{FIG-convfour}.

It is not obvious that this holds for the chaotic regime as well: there,
$x$ keeps changing on all scales, not just increasingly small ones, and
multiplication induces a folding of small and large scales. 
Interestingly, however, similar structures can be found in the chaotic regime as
well: even though a 3D plot of the CR versus time looks unstructured and
chaotic, a Fourier transform averaged over sufficiently many
initial conditions often reveals diagonal stripes (Fig. \ref{FIG-convfour},
bottom),  this time tilted  in the opposite way
-- which indicates information traveling to coarser scales rather
than finer ones. Although the structures are less clear than for convergence,
an approximate correspondence between the slope of the structures and the Lyapunov 
exponent still holds for many values of $a$, 
as shown by the diagonal lines in the plots. In other cases, the structures
are less clear:
in particular, for $a=4.0$ the plot shows strong diagonal structures with a slope of
2 overlaid on a weak diagonal with a  slope of approximately $1$.
The latter is expected from the value of the Lyapunov exponent.

It should be noted that using the binary representation instead of the
CR yield similar pictures (albeit more noisy) for the convergence to a
fixed point, but shows no discernible structures for the convergence to the
chaotic attractor.

\section{Summary}
\label{SEC-summary}
We have presented a study of the loss of information through the nonlinear
dynamics of the logistic map, using analytical means for short times and
numerics and heuristic arguments for long times. 
As Secs. \ref{SEC-info} and \ref{SEC-commun} have shown,
 different processes are
relevant in the different regimes:
in the small-$a$ regime, shrinking phase space quickly makes postdiction 
impossible and prediction trivial, thus eliminating the possibility of 
communication. 
In the chaotic regime (very high $a$), phase space is largely conserved, but
sensitivity to initial conditions prevents both
prediction and postdiction for longer times. 
The bifurcation regime provides a middle
ground: some information about the initial state persists, determining what
branch of the cycle one finds the system in. The basins of attraction for
the branches have a fractal structure, which means that some large
intervals of initial values exist that lead to one branch with certainty.
At the bifurcation points, convergence to the final state is slow, which
makes some transfer of information possible for intermediate times.

The clockwork representation introduced in Sec. \ref{SEC-rep} is a
continuous generalization of the usual discrete (binary, decimal etc.)
representations in which 
addition and multiplication of two objects are more transparent 
than in the discrete case, and which allows for a more elegant
visualization of the flow of information between scales and the convergence
to fixed points. It is even possible to identify the slope of structures in
the Fourier transform of the CR with the Lyapunov exponent of the map.
While the CR thus seems to be a conceptual and visual tool of some use,
 it remains to be seen whether this representation 
will find additional applications in analyzing dynamical systems.

\begin{acknowledgments}
This work is partially supported by NSF through grant No. DMS-0083885.
M.K. is supported by NSF through grant No. DMR-01-18213.
\end{acknowledgments}

%\bibliography{bibliography}

\begin{thebibliography}{17}
\bibitem{endnote18}{This can be regarded as defining what the observed quantity is. Transformations commonly used to switch from one set of variables/equations to another do not conserve this property.}
\end{thebibliography}

\end{document}